\documentclass[12pt]{article}
\usepackage{epsfig}
\usepackage{comment}
\usepackage{latexsym}
\usepackage{color}
\newcommand{\mysquare}[0]{\raise-.2ex\hbox{{\Large$\Box$}}}
\def\lsim{\mathrel{\rlap {\raise.5ex\hbox{$ < $}}
{\lower.5ex\hbox{$\sim$}}}}
\def\gsim{\mathrel{\rlap {\raise.5ex\hbox{$ > $}}
{\lower.5ex\hbox{$\sim$}}}} \topmargin -1.5cm \textheight=22.5cm
\catcode`\@=11
\newcount\hour
\newcount\minute
\newtoks\amorpm
\hour=\time\divide\hour by60 \minute=\time{\multiply\hour by60
\global\advance\minute by-\hour}
\edef\standardtime{{\ifnum\hour<12 \global\amorpm={am}%
        \else\global\amorpm={pm}\advance\hour by-12 \fi
        \ifnum\hour=0 \hour=12 \fi
        \number\hour:\ifnum\minute<10 0\fi\number\minute\the\amorpm}}
\edef\militarytime{\number\hour:\ifnum\minute<10 0\fi\number\minute}
\def\draftlabel#1{{\@bsphack\if@filesw {\let\thepage\relax
   \xdef\@gtempa{\write\@auxout{\string
      \newlabel{#1}{{\@currentlabel}{\thepage}}}}}\@gtempa
   \if@nobreak \ifvmode\nobreak\fi\fi\fi\@esphack}
        \gdef\@eqnlabel{#1}}
\def\@eqnlabel{}
\def\@vacuum{}
\def\draftmarginnote#1{\marginpar{\raggedright\scriptsize\tt#1}}
\def\draft{\oddsidemargin -.2truein
        \def\@oddfoot{\sl preliminary draft \hfil
        \rm\thepage\hfil\sl\today\quad\militarytime}
        \let\@evenfoot\@oddfoot \overfullrule 3pt
        \let\label=\draftlabel
        \let\marginnote=\draftmarginnote
   \def\@eqnnum{(\theequation)\rlap{\k

 ern\marginparsep\tt\@eqnlabel}%
\global\let\@eqnlabel\@vacuum}  }

\newcommand{\be}[0]{\begin{equation}}
\newcommand{\ee}[0]{\end{equation}}
\newcommand{\ba}[0]{\begin{eqnarray}}
\newcommand{\ea}[0]{\end{eqnarray}}

\def\bs{\begin{subequations}}
\def\es{\end{subequations}}

\def\thebibliography#1{%
\vskip 0.5cm \centerline{\bf \Large References}
\list{%
[\arabic{enumi}]}{\settowidth\labelwidth{[#1]}
\leftmargin\labelwidth \advance\leftmargin\labelsep
\usecounter{enumi}}
\def\newblock{\hskip .11em plus .33em minus .07em}
\sloppy\clubpenalty4000\widowpenalty4000 \sfcode`\.=1000\relax}

\renewcommand{\theequation}{\arabic{section}.\arabic{equation}}

\renewcommand{\section}{\setcounter{equation}{0}\@startsection
{section}{1}{0mm}{-\baselineskip}{0.5\baselineskip}
{\normalfont\Large\bfseries}}

\renewcommand{\subsection}{\@startsection
{subsection}{2}{0mm}{-\baselineskip}{0.5\baselineskip}
{\normalfont\large\bfseries}}

\renewcommand{\subsubsection}{\@startsection
{subsubsection}{3}{0mm}{-\baselineskip}{0.5\baselineskip}
{\normalfont\normalsize\slshape}}

\usepackage{amssymb,amsfonts}
\usepackage{graphicx}
\usepackage{cite}

\def\bc{\begin{center}}
\def\ec{\end{center}}
\def\bea{\begin{eqnarray}}
\def\eea{\end{eqnarray}}

\def\k{\kappa}

\def\tt{\tilde t}

\def\and{\quad\mbox{and}\quad}

\def\k{\kappa}

\def\ov{\overline}

\renewcommand{\@}[1]{\sqrt{#1}}

\newcommand{\eq}[1]{(\ref{#1})}

\def\ffract#1#2{\raise .35 em\hbox{$\scriptstyle#1$}\kern-.25em/
\kern-.2em\lower .22 em \hbox{$\scriptstyle#2$}}

\newcommand{\qbin}[2]{\left[^{#1}_{#2}\right]}

\begin{document}
\begin{titlepage}
\begin{flushright} LPTENS--06/57
\end{flushright}

\vspace{2mm}

\begin{centering}
{\bf\huge Magic $N=2$ supergravities from }\\
\vspace{2mm}
{\bf\huge hyper-free superstrings$^\ast$}\\

\vspace{1cm}
 {\Large Yacine Dolivet, Bernard Julia and Costas Kounnas}

\vspace{4mm} Laboratoire de Physique Th\'eorique,
Ecole Normale Sup\'erieure,$^\dagger$ \\
24 rue Lhomond, F--75231 Paris Cedex 05, France\\
\vskip 4mm {\em  bjulia-AT-lpt.ens.fr,
kounnas-AT-lpt.ens.fr}

\vskip 2cm

{\bf\Large Abstract}

\end{centering}

\noindent We show by explicit construction the existence of various
four dimensional models of type II superstrings with $N=2$
supersymmetry, purely vector multiplet spectrum and no
hypermultiplets. Among these, two are of special interest, at the
field theory level they correspond to the two exceptional $N=2$
supergravities of the magic square that have the same massless scalar
field content as pure $N=6$ supergravity and  $N=3$  supergravity coupled
to three extra vector multiplets. The $N=2$  model of the magic  square that
is associated to $N=6$ supergravity is  very peculiar since not only
the scalar degrees of freedom but all the bosonic massless degrees
of freedom are the same in both theories.
 All presented hyper-free  $N=2$
models are based on asymmetric orbifold constructions with ${\rm
\cal N}=(4,1)$ world-sheet superconformal symmetry and utilize the 2d
fermionic construction techniques. The two exceptional  $N=2$ models
of the magic square are constructed  via  a ``twisting mechanism"
that eliminates the extra gravitini of the $N=6$  and $N=3$ extended
supergravities and creates at the same time the extra spin-${1\over 2}$
fermions and spin-1 gauge bosons which are necessary to balance the
numbers of bosons and fermions. Theories of the magic square with the same 
amount of supersymmetry in three and five space-time dimensions are  constructed as well,
via stringy reduction and oxidation from the corresponding four-dimensional models.


\vspace{5pt} \vfill \hrule width 6.7cm \vskip.1mm{\small \small
\small \noindent $^\ast$\ Research partially supported by the EU
(under the contracts MRTN-CT-2004-005104, MRTN-CT-2004-512194,
MRTN-CT-2004-503369, MEXT-CT-2003-509661), CNRS PICS 2530, 3059 and
3747, ANR (CNRS-USAR) contract
 05-BLAN-0079-01.\\
$^\dagger$\ Unit{\'e} mixte  du CNRS et de l'Ecole Normale
Sup{\'e}rieure associ\'ee \`a l'universit\'e Pierre et Marie Curie
(Paris 6), UMR 8549.}

\end{titlepage}

\newpage
\setcounter{footnote}{0}
\renewcommand{\thefootnote}{\arabic{footnote}}
 \setlength{\baselineskip}{.7cm} \setlength{\parskip}{.2cm}


\setcounter{section}{0}
\section{Introduction}

\noindent In four dimensions there is an exceptional family  of $N=2$ supergravities
\cite{gst} which are known to be in correspondence with the
symmetric spaces of the ``magic square" of Freudenthal-Rozenfeld-Tits
\cite{frt}. Two of the four $N=2$ exceptional theories in four
dimensions are associated to the Jordan algebras $J_3^{\mathbb{C}}$
and $J_3^{\mathbb{H}}$. They have the additional remarkable property
that they share the same scalar field contents as some supergravity
models with more supersymmetry. Their scalar manifolds
\begin{equation}
{\rm {\cal M}}_3={ SU(3,3)\over S( U(3) \times U(3))}, ~~~~~~ {\rm
{\cal M} }_6={SO^{*}(12)\over U(6)}
\end{equation}
appear also in $N=3$ supergravity with 3 vector supermultiplets and
in (pure) $N=6$ supergravity, respectively.

\noindent The $N=2$ model which is associated to $N=6$ supergravity
appears to be most peculiar. Not only the scalar fields of both theories
are the same but 
all the other bosonic degrees of freedom are also the same. This
remarkable property motivated us to realize this exceptional $N=2$
theory at the superstring level. In order to proceed in this
direction we found useful to focus our attention more generally on
the construction of $N=2$ theories with purely vector multiplet
spectrum i.e. without hypermultiplets. As a consequence, the scalar
fields will all belong to vector multiplets and parameterize a
projective special K\"ahler manifold
\cite{VanWit},\cite{Cremmer:1984hj}, \cite{cvp}. All theories of the
``magic square" belong to the class of hyper-free theories.

\noindent In this work we construct four dimensional type II
superstring backgrounds with $N=2$ space-time supersymmetry
\footnote{Here and in the following $N=N_4$ corresponds to the
number of space-time supersymmetries with respect to the four
dimensional super-Poincar\'e algebra. It will sometimes be written as
$N=N_L+N_R$ to recall the world-sheet chirality responsible for the
supercharges}. The  $N=2$ hyper-free models are constructed
via asymmetric orbifolds utilizing the free 2d-fermion construction
\cite{abk,klt}. Due to the left-right asymmetric construction the
universal axion-dilaton pair, $S$, will not be part of a
hypermultiplet, as would be the case for type II geometric
compactifications on a Calabi-Yau manifold. Instead, $S$ will belong
to a vector multiplet as in the heterotic-like compactifications.
This is linked to the fact that supersymmetry is realized in an
asymmetric way from the world-sheet point of view \cite{abk,fk, vw},
as in the heterotic case or in the  asymmetric constructions of type
II obtained in  \cite{sv, gk}. Instead of having ${\rm\cal
N}=(2,2)$ superconformal models as is the case for a Calabi-Yau
compactification we rather have ${\rm \cal N}=(4,1)$ superconformal
symmetry on the world-sheet \cite{bd}. We may note that the c-map
correspondence exchanges hypermultiplets and vector multiplets, it
is a $T$ duality in 3d conjugated by scalar/vector dualities, we shall
return to this in due course.

\noindent In section \ref{s:minimal} we start with the construction
of the ``$S-$minimal'' $N=2$ type II superstring model which
contains a single minimally coupled vector multiplet,  $S$,
associated to the axion-dilaton pair. The $S-$minimal theory is
universal in the sense that it will be part of the spectrum of all
the other more complex hyper-free models.

\noindent In section \ref{s:magicsquare} we recall the properties of
some exceptional $N=2$ supergravities \cite{gst} which are known to
be in correspondence with the projective special K\"ahler symmetric
spaces of the ``magic square" of Freudenthal-Rozenfeld-Tits
\cite{frt}.

\noindent In section \ref{s:magicsquareN=6} and
\ref{s:magicsquareN=3} we construct as type II superstrings the two
$N=2$ theories of the ``magic square" associated to the $N=6$ and
$N=3$ supergravity theories respectively. We first recall how to
realize the associated theories with higher supersymmetry ($N=6$ and
$N=3$) and then we introduce a new mechanism which  reduces
supersymmetry \emph{without} changing the scalar content of the
models. Its sole effect will be to somehow replace the gravitini
that we want to get rid off,  by spin-$1/2$ fermions and spin-$1$
gauge bosons, thereby giving us precisely the spectrum of the $N=2$
theories we are looking for.

\noindent Two extra theories of the magic square in three space-time
dimensions, ${\rm \cal M}^{D=3}_6={E_{7(-5)}\over SU(2) \times
SO(12)}~,$ and $~{\rm \cal M}^{D=3}_3={E_{6(2)}\over SU(2)\times
SU(6)}$ are constructed in section \ref{s:magicsquareN=6} and
\ref{s:magicsquareN=3} by stringy dimensional reduction from the
corresponding four dimensional magic theories. Furthermore, by
oxidation in five space-time dimensions the construction of ${\rm
\cal M}^{D=5}_6={SU^*(6)\over USP(6)}~$ of the magic square is also
obtained. In our stringy set-up, there is an obstruction to define
${\rm \cal M}^{D=5}_3 = {SL(3,C)\over SU(3)}~$ since all six
right-moving coordinates are twisted which prevent the oxidation
procedure.

\noindent Section \ref{s:conclusions} summarizes our results.

\section{The minimal hyper-free theory}
\label{s:minimal}
 \noindent Our first theory is a type IIA four dimensional $N=2$
 model which contains the
graviton supermultiplet and one additional vector multiplet in its
massless spectrum. The vector multiplet consists of the universal
dilaton and axion fields. The model can be easily obtained via the
fermionic construction \cite{abk}, \cite{klt}, \cite{fk}; the basis
of sets $\{F,S,\ov{S},b_1,\ov{b}_1,\ov{ b}_2,\ov{b}_3\}$ is the main data relevant
for the construction of the corresponding model which we call
$\langle F,S,\ov{S},b_1,\ov{b}_1,\ov{ b}_2,\ov{b}_3 \rangle$ but it implies
also a choice of some generalized GSO projection coefficients (GGSO or discrete
torsion). Here $F$ is the set containing all the
fermions of the model in the light-cone gauge, namely two
world-sheet fermions $\psi^\mu$, the 12 fermionized internal coordinates
$\{y^I,w^I\}$ with $\partial X^I=y^Iw^I, I=1\ldots 6$, their 6
world-sheet supersymmetric partners $\chi^I$ as well as all the
corresponding right- moving 2d fermions. The sets $S$, $\ov{S}$ that
define the left- and right- supersymmetric GSO projections are given
by\cite{abk}:
\begin{equation}
S=\{\psi^\mu,\chi^{1,\ldots, 6}\}, ~~~~~~~\ov{S}=
\{\ov{\psi}^\mu,\ov{\chi}^{1,\ldots, 6}\}~.
\end{equation}
The model defined by the sets $\{F,S,\ov{S}\}$ gives rise to the
usual $N=4+4$ supersymmetric background. In order to reduce the
left- plus right-moving space-time supersymmetry from $N=4+4$ to $N=2+1$ resp. to
$N=2+0$, it is necessary to include some extra ``supersymmetry
breaking sets", $b_1,\ov{b}_1,\ov{ b}_2 $ resp. $b_1,\ov{b}_1,\ov{ b}_2,\ov{b}_3 $
that define left-right-asymmetric projections of the type:
$$
\left(Z_2\right)_{\rm left}\times
\left( Z_2\times Z_2\right)_{\rm right}
$$
resp.
$$
\left(Z_2\right)_{\rm left}\times
\left( Z_2\times Z_2\times Z_2\right)_{\rm right}
$$
with
$$
~{b}_1 = \{~{\psi}^\mu,~{\chi}^{1,2},
~{y}^{3,4},~{y}^{5,6},~{y}^1,{w}^1\,|\,\ov{y}^5,\ov{w}^5\,\}
$$
$$
~~\ov{b}_1 = \{~\ov{\psi}^\mu,~\ov{\chi}^{1,2},~\ov{y}^{3,4},~\ov{y}^{5,6},
 ~\ov{y}^1,~\ov{w}^1 \,|\,
{y}^5,{w}^5 \}
$$
$$
~\ov{b}_2 = \{~\ov{\psi}^\mu,~\ov{\chi}^{3,4},~\ov{y}^{1,2},~\ov{w}^{5,6},
~\ov{y}^3,~\ov{w}^3 \,|\,
{y}^6,{w}^6\}
$$
\begin{equation}
~\ov{b}_3 = \{~\ov{\psi}^\mu,~\ov{\chi}^{5,6},~\ov{w}^{1,2},~\ov{w}^{3,4},
 ~\ov{y}^6,~\ov{ w}^6
\,|\,{y}^2,{w}^2 \}
\end{equation}
The above choices of the ``supersymmetry breaking sets" define
consistent models since they satisfy all the overlapping
conditions necessary in the fermionic construction \cite{abk}.
$$
~\ov{b}_1\cap~\ov{ b}_2=\{~\ov{\psi}^\mu,~\ov{y}^{1,3}\} ;~~~~
 ~\ov{ b}_1\cap {b}_1=\{~\ov{y}^5\,|\,
{y}^5\};~~~~ ~\ov{b}_2\cap {b_1}=\{~\ov{w}^5\,|\, {y}^6\}
$$
\begin{equation}
 ~\ov{b}_3\cap~\ov{ b}_1=\{~\ov{\psi}^\mu,~\ov{y}^6,~\ov{w}^1 \};
~~~~~\ov{b}_3\cap ~\ov{b}_2=\{~\ov{\psi}^\mu,
~\ov{w}^{3,6}\};~~~~~\ov{b}_3\cap {b}_1=\emptyset
\end{equation}
and \begin{equation}
~\ov{b}_1\cap ~\ov{b}_2\cap ~\ov{b}_3\cap {b}_1=\emptyset.
\end{equation}
The presence of some left- and right-fermionized coordinates,
$\{y_i,w_i|\ov{y}_j,\ov{w}_j\}$, in the $~\ov{b}_I,I=1,2,3$ and ${b}_1$
guarantees the free-action of the  $Z_2$'s defining the
asymmetric orbifolds \cite{nsv} and thus the absence
of massless states coming
from the ``twisted sectors". The models are  defined completely once
we specify the signs of the generalized GSO coefficients (GGSO). We
have chosen to take minus one for all  but one:
\begin{equation}(-1)^{~\ov{b}_1}=(-1)^{~\ov{b}_2}=
(-1)^{~\ov{b}_3}=(-1)^{{b}_1}=-1\, ,
\end{equation} in
addition to the choices \begin{equation}(-1)^F=1,~~~~
(-1)^S=(-1)^{\ov{S}}=-1 \, ,\end{equation}
 that define the usual $N=4+4$ model in type IIA theory.

\noindent The model $\langle F,S,\ov{S},b_1,\ov{b}_1,\ov{ b}_2
\rangle$ without the set $b_3$, defines a theory possessing $N=2+1$
space-time supersymmetry and $SU(3,1)$ as a non-compact group of
global symmetry \cite{fk}. The additional set $\ov{b}_3$ enables us
to define the  model $\langle F,S,\ov{S},b_1,\ov{b}_1,\ov{
b}_2,\ov{b}_3  \rangle$ in which  all left- moving space-time
supersymmetries are broken. This results in a ${\rm \cal N}=(4,1)$
superconformal theory on the world-sheet and  $N=2+0$ space-time
supersymmetry.

\noindent Since $\ov{b}_3$ defines a $Z_2$ action that acts freely,
it does not introduce additional states from its twisted sector in
the massless spectrum either. Besides, all the bosonic fields from the
``parent'' $N=2+1$ theory are projected out apart from the graviton,
the dilaton-axion pair and two vector gauge fields. One is the
graviphoton of the gravity multiplet and the other corresponds to a
matter vector multiplet.

\noindent Although the massless spectrum of the $N=2$ model consists
of the gravitational multiplet and just one vector multiplet, it is
important to determine the coupling between the vector and scalar
fields and the structure of the scalar moduli space. It is well
known  \cite{fk}
 that the axion-dilaton pair parameterizes a coset space
which is topologically a pseudosphere
\begin{equation}
\frac{SU(1,1)}{U(1)}~.
\end{equation}
Let us recall that the $N=2$ effective
supergravity allows a priori two types of couplings between the vector and
scalar fields which are distinguished by the \emph{curvature} of  the
moduli space \cite{fk},\cite{cvp}  even though their topological
structure is the same. Indeed, when the coupling between the scalar
and vector fields is {\rm ``non-minimal"}  the K\"ahler potential is
given by
\begin{equation}
K=-3\log (S-\ov{S})~.
\end{equation}
On the other hand, when the coupling is {\rm ``minimal"} the
K\"ahler potential is given by
\begin{equation}
K=-\log(S-\ov{S})~.
\end{equation}
From the form of the axion-dilaton kinetic terms in type II
superstrings we know that this last case applies to the structure of
the axion-dilaton moduli space \cite{fk}.
The Poincar\'e half plane is always 
 isometrically embedded in a possibly bigger moduli space $G/H$.

\noindent In the language of $N=2$ supergravity we recall that $S$
corresponds to a non-homogeneous coordinate on the moduli space
\cite{VanWit},\cite{Cremmer:1984hj},\cite{cvp}. Homogeneous
coordinates are introduced through $S=\frac{Z}{Z_0}$ and more
generally $t_i=\frac{Z_i}{Z_0}$, $i=1\ldots n$ if there are $2n+2$
moduli of the vector multiplets. In terms of these, the K\"ahler
potential of vector-moduli space  is fixed by the prepotential
$F(Z_0,Z_i)$ - which is a holomorphic homogeneous function of degree
2 - through\cite{VanWit},\cite{Cremmer:1984hj},\cite{cvp}
\begin{equation}
K=-\log ~i{\rm Im}[~\ov{Z}_I \partial^I F(Z_J)~]~,
\end{equation}
with the index $I=(0,i)$. It can alternatively be expressed in terms of the
non-homogeneous coordinates $t_i$ and the function $f(t_i)$ defined
by $F(Z_I)=-i Z_0^2f(t_i)$ through\footnote{In this change of
notation one makes use of the invariance of the theory under a
K\"ahler transform of the K\"ahler potential
\begin{equation}
 K\rightarrow K+\Lambda+\ov{\Lambda}
\end{equation}
with $\Lambda$ an arbitrary holomorphic function of the moduli.}
\begin{equation}
K=-\log ~[~ 2(f(t_i)+\ov{f}(\ov{t}_i))-(t_i-\ov{t}_i) (\partial^i
f-\ov{\partial}^i\ov{f})~]~.
\end{equation}
In the case of non-minimal coupling with coset space $SU(1,1)/U(1)$
the prepotential is cubic in $Z$ and given by $F(Z,Z_0)=i\frac{Z^3}{Z_0}$.
On the other hand, the prepotential in the case of minimal coupling
to the vector fields is $F(Z,Z_0)=Z^2-Z_0^2$. The latter
prepotential is the one we must use to describe the axion-dilaton
pair in the case of interest. Therefore, we have shown that it is
possible to construct a ``minimally" coupled $N=2$
hyper-free model whose scalar sector forms a special
K\"ahler coset space
$$
\frac{SU(1,1)}{U(1)}~.
$$
This is interesting and rather unexpected. Indeed, in the spirit
of the c-map \cite{cfg} the distinction between the ``minimal" and
``non-minimal" structures is equivalently expressed by saying that
the dimensional reduction to three space-time dimensions of these
models on a circle, gives rise to 3d supergravity models with
different scalar manifolds (after dualization in three dimensions of
the vector fields into scalars). In the case of a
``non-minimal" coupling one finds that the scalars together with the
scalar-dual of the vector gauge field, parameterize the special
quaternionic manifold\footnote{One can read \cite{vpro} for a
discussion of special K\"ahler and special quaternionic
manifolds.}
\begin{equation}
\frac{G_{2(2)}}{SO(4)}~,
\end{equation}
whereas in the case of ``minimal" coupling one finds that the scalar
parameterize the special  quaternionic-K\"ahler manifold,
\begin{equation}
\frac{U(2,2)}{U(2)\times U(2)}~,
\end{equation}
which is somewhat exotic from the point of view of Calabi-Yau or
symmetric orbifold ${\rm \cal N}=(2,2)$ compactifications.

\section{Hyper-free $N=2$ theories of the ``magic square"}
\label{s:magicsquare}

\noindent More general hyper-free $N=2$ models can be constructed
with higher number $n_V$ of vector multiplets via asymmetric
orbifold constructions starting from  type II superstrings  with
$N=2+4$,  $N=2+1$ or eventually $N=2+2$ initial supersymmetry. The
breaking of the right-moving supersymmetry via freely acting
asymmetric orbifold gives rise to $N=2+0$ hyper-free models. In all
constructions of this type, the axion-dilaton pair belongs to one of the
vector multiplets and it appears always with ``minimal
coupling". Since in our construction the last projection is assumed
to act freely, the scalar K\"ahler manifold of the final $N=2$
theory is necessarily a sub-manifold of ${\rm \cal M}^{N=2}$ the
scalar manifold of the initial
``mother" theory with higher supersymmetry, let us consider for instance
for N=4+2 or N=2+1:
\begin{equation}
{\rm \cal M}^{N=2} \subseteq{SO^{*}(12)\over U(6)},~~~~{\rm
or}~~~~{\rm \cal M}^{N=2} \subseteq{ SU(3,3)\over S( U(3) \times
U(3))}.
\end{equation}
The special cases where the dimension of the scalar manifold is
maximal, dim${\rm \cal M}^{N=2}_6$=30 and dim${\rm \cal
M}^{N=2}_3$=18,  hold for $N=2$ theories which are known to be in
correspondence with the projective special K\"ahler symmetric spaces
of the ``magic square" of Freudenthal-Rozenfeld-Tits \cite{frt},
\cite{gst}.

\noindent These two $N=2$ ``magic" theories are associated to
 the Jordan algebras $J_3^{\mathbb{H}}$
and $J_3^{\mathbb{C}}$. They are known to have the same scalar field
content as some supergravity models with more supersymmetries. The
$N=2$ theory that is associated to  $N=6$ supergravity appears
to be the most peculiar one. Not only the scalar fields but also the
gauge bosons degrees of freedom are the same.

\noindent In the next two sections we provide a construction of  the
two hyper-free $N=2$ ``magic" theories at the superstring level.

\noindent Before that, let us  recall some properties of the $N=6$
supergravity, namely the helicity structure of the $N=6$ graviton
($\mathbf{G}$) and gravitino ($\mathbf{g}$) supermultiplets
$$
\mathbf{G}  : (+2,+\frac{3}{2}^6,+1^{15},+
\frac{1}{2}^{20},0^{15},-\frac{1}{2}^6,-1)\oplus
(+1,+\frac{1}{2}^6,0^{15},-\frac{1}{2}^{20}, -1^{15},
-\frac{3}{2}^6,-2)
$$
\begin{equation}
\mathbf{g} : (+\frac{3}{2},+1^{6},+\frac{1}{2}^{15},0^{20},
-\frac{1}{2}^{15},-1^{6},-\frac{3}{2})
\end{equation}

\noindent
where we indicate the multiplicity of each helicity in exponent. The
branching rule under $N=6\rightarrow N=2$,
\begin{equation}
\label{eq:branching62} \mathbf{G}\rightarrow \mathbf{G}\oplus
4\mathbf{g} \oplus 7\mathbf{V} \oplus 4\mathbf{H}
\end{equation}
where $\mathbf{V}$ and $\mathbf{H}$ denote respectively
vector-multiplets and hyper-multiplets. The 30 scalars of the $N=6$
supergravity parameterize a coset
\begin{equation}
\frac{SO^*(12)}{U(6)}~.
\end{equation}
Remarkably, one of the $N=2$ theories of the magic square\cite{gst}
 possesses exactly the same
bosonic content as the $N=6$ (pure) supergravity. It is associated
to the Jordan algebra called $J_3^{\mathbb{H}}$. Both $N=6$ and
$N=2$ supergravities contain one graviton, 15 vector fields (one of
them being the graviphoton and 14 belonging to vector multiplets in
the case of the $N=2$) and the 30 scalars with the structure we have
just indicated.

\noindent Before we proceed further, we would like to make some
remarks concerning the fermionic spectrum : from the decomposition
of the eq.\eq{eq:branching62}, $N=6$ supergravity contains 26 extra
fermions in addition to the 2 gravitini from the graviton multiplet
$G|_{N=2}$, namely 22 have spin $1/2$ and the last 4 are spin-$3/2$
gravitini. This is the same as the number of fermionic degrees of
freedom of the $J_3^{\mathbb{H}}$ supergravity.
Hence the difference is that the extra 4
gravitini must be replaced in the $N=2$ model by 4 spin-$1/2$
fermions.

\section{Superstring construction of the magic
${\rm \cal M}^{N=2}_6$}\label{s:magicsquareN=6}

\noindent This section is devoted to the superstring construction of
the exceptional $N=2$ supergravity based on the Jordan algebra
$J_3^{\mathbb{H}}$ \cite{gst} which contains at the massless
level the same number of scalars and gauge bosons as pure $N=6$
supergravity.

\noindent The simplest way to get a $N=6$ theory in the type IIA
setup is to start from the type II superstring with $N=4+4$
supersymmetry and then to use a freely acting asymmetric $Z_2$
orbifold which reduces the left-moving supersymmetries to
$N=2+4$\cite{fk}, \cite{sv}, \cite{gk}. In the fermionic construction language\cite{abk},
we start with the $N=4+4$ model $\langle F,S,\ov{S}\rangle$. Then,
the $Z_2$ asymmetric breaking to $N=2+4$ is obtained by
choosing one additional supersymmetry breaking set for instance
$b'$ \cite{fk},
\begin{equation}
 b'=\{\psi^\mu,
\chi^{1,2},y^{3,4,5,6},y^1,w^1\, |\,\ov{y}^1,\ov{w}^1 \},
\end{equation}
and fixing the GGSO projection by the choice of sign $(-1)^{b'}=-1$.\\
 Although the $\langle F,S,\ov{S},b'\rangle$ model
 defines an initial $N=2+4$ theory, it will turn out that it is not
 a good starting point to reduce the right-moving gravitini
 and obtain the magic $N=2+0$ theory.

\subsection{A failed attempt}
 Indeed, we could add an additional breaking set $\ov{b}_{\ov{S}}$
\begin{equation}
 \ov{b}_{\ov{S}}= \{y^2,w^2\, |\,\ov{y}^2,\ov{w}^2 \}\cup \ov{S}
\end{equation}
that would break $N=2+4$ to $N=2+0$ removing the four right- moving
gravitini.  However, this would remove at the same time the RR-scalars and
so, the number of remaining scalar degrees of freedom of the $N=2$
theory would be 14 instead of 30. The scalar manifold of this hyper-free
$N=2$ model, $\langle F,S,\ov{S},b' , \ov{b}_{\ov{S}}\rangle$,
\begin{equation}
{\rm \cal M}^{N=2}={SU(1,1)\over U(1)}\times {SO(2,6)\over
SO(2)\times SO(6)}~\subset~ \frac{SO^*(12)}{U(6)}
\end{equation}
 is a sub-manifold of the desired ${\rm \cal M}^{N=2}_6$
 with dimension $14$ instead of  $30$.

 \noindent To obtain the desired
 ``magic" $N=2+0$ it seems necessary
 to construct in a different way the initial $N=2+4$ theory so that,
the RR scalars and gauge bosons  survive the additional
supersymmetry breaking projection from $N=2+4$ to $N=2+0$.

\subsection{$N=(2+2) +4$ construction with
$(3/2)\leftrightarrow(1/2)$ magic twist}

\noindent To define the asymmetric orbifold  mechanism which has the
property to replace the 4 gravitini of the $N=6$ by 4 fermions and
at the same time keeps the RR-scalars, it is necessary to
prepare the initial $N=2+4$ theory in a more sophisticated way
that we will describe in detail below.

\noindent We start with the string model $\langle
F,S,\ov{S},b\rangle$, where $b$ is {\em purely left-moving}
\begin{equation}
b=\{\psi^\mu, \chi^{1,2},y^{3,4,5,6}\}.
\end{equation}
$b$ defines an asymmetric, non-freely acting orbifold  acting on
$\langle F,S,\ov{S}\rangle$ model and breaks the supersymmetry (in
the untwisted sector), from $N=4+4 \rightarrow N=2+4$. The twisting
action of $b$ on the untwisted sector is thus similar to the one of
$b'$. There is however a \emph{fundamental difference} between $b$
and $b'$; in $b$  all elements are left-moving. This holomorphic
property will be crucial  in what follows.

\noindent Different choices of GGSO projection give different
models. Choosing for instance $(-1)^{b}=-1$, the massless spectrum
would be given by:

 \noindent
\underline{$\hat{\emptyset}$ super-sector} ~
$$
\left[ (\psi^\mu,\chi^{1,2}) \, \oplus\, sp(\psi^\mu,\chi^{1,2})_-
sp(\chi^{3,4,5,6})_+ \right]
$$
\begin{equation}
\otimes \left[ (\ov{\psi}^\mu,\ov{\chi}^{1,\ldots,6}) \, \oplus\,
sp(\ov{\psi}^\mu,\ov{\chi}^{1,\ldots,6})_-\right],
\end{equation}

\noindent \underline{$\hat{b}$ super-sector} ~
$$
\left[ sp(\psi^\mu,\chi^{1,2})_-sp(y^{3,4,5,6})_+\, \oplus \,
sp(\chi^{3,4,5,6})_- sp(y^{3,4,5,6})_+ \right]
$$
 \begin{equation}
  \otimes \left[ (\ov{\psi}^\mu,\ov{\chi}^{1,\ldots,6})
\, \oplus\, sp(\ov{\psi}^\mu,\ov{\chi}^{1,\ldots,6})_-\right].
\end{equation}

\noindent In each super-sector our notation indicates the massless
states in the sectors $\left[ NS \, \oplus \, R \right]\otimes\left[
\ov{NS} \, \oplus\, \ov{R} \right] $. Even though the untwisted
{$\hat{\emptyset}$ super-sector} corresponds to a theory with $N=2+4$
supersymmetry as in the previous $b'$-construction, here the
supersymmetry is extended to $N=(2+2)+4$ because now the twisted
super-sector $\hat{b}$ contains massless states and among them {\it
two extra left-moving gravitini}. This supersymmetric extensions
happens due to the left-moving ``holomorphic structure'' of $b$.
Thus, the $\langle F,S,\ov{S}, b\rangle $ twisted construction still
has maximal $N=8$ supersymmetry constructed in a $b$-twisted manner.
We should stress here that this supersymmetric extension from
$N=6\rightarrow N=8$ of the $\langle F,S,\ov{S}, b\rangle $ model
{\it is a stringy phenomenon} and can be seen algebraically at the
level of the modular invariant partition function which implies the
inclusion in the spectrum of the $b$-twisted sectors with $h\ne 0$:

$$
Z_b=\frac{1}{|\eta|^{8}}\frac{1}{4}\sum_{h,g}
Z_{6,6}\left[^h_g\right]|_{SO(6)} \sum_{a,b}(-1)^{a+b+ab}\,
\theta\qbin{a+h}{b+g}\theta\qbin{a-h}{b-g}\theta\qbin{a}{b}^2
$$
\begin{equation}
\times \frac{1}{2}\sum_{\ov{a},\ov{b}}(-1)^{\ov{a}+
\ov{b}+\ov{a}\ov{b}}\,\ov{\theta}\qbin{\ov{a}}{\ov{b}}^4
\end{equation}
where $\left.Z_{6,6}\qbin{h}{g}\right|_{SO(6)}$ is the contribution
of the six internal coordinates
$\partial\phi_L^{I}=(y\omega)^I,~I=1,2,...6$. The directions
$\phi_L^{3,4,5,6}$ are twisted by $Z_2$ induced by $b$. Remember
that in the fermionic construction the coordinate currents are given
in terms of the two dimensional free fermions, so that the $Z_2$
acts on $y^{3,4,5,6}$ only. $\omega^{3,4,5,6}$, 
$y^{1,2} $ and $ \omega^{1,2}$ are invariant under $Z_2$.
$$
Z_{6,6}\qbin{h}{g}_{SO(6)} ={1\over 2|\eta|^{4}}~
\sum_{\gamma,\delta}\theta\qbin{\gamma}{\delta}^2_{y^{1,2},\omega^{1,2}}~
\ov{\theta}\qbin{\gamma}{\delta}^2_{\ov{y}^{1,2},\ov{\omega}^{1,2}}~
$$
\begin{equation}
\times~{(-1)^{\gamma g+\delta h}\over
|\eta|^8}~\theta\qbin{\gamma}{\delta}^2_{\omega^{3,4,5,6}}
\theta\qbin{\gamma+h}{\delta+g}_{y^{3,4}}\theta\qbin{\gamma-h}
{\delta-g}_{y^{5,6}}~\ov{\theta}\qbin{\gamma}{\delta}^4_{\ov{y}^{3,4,5,6},
\ov{\omega}^{3,4,5,6}}.
\end{equation}

\noindent To see that this leads to an $N=4+4$ supersymmetry we
first perform the sum over the $(a,b)$ indices using the Jacobi
identity
$$
\frac{1}{2}\sum_{a,b}(-1)^{a+b+ab}~\theta\qbin{a+h}{b+g}(v)~
\theta\qbin{a-h}{b-g}(v)~\theta\qbin{a}{b}^2(v)$$
\begin{equation}
=-\theta\qbin{1}{1}^2(v)~\theta\qbin{1+h}{1+g}(v)~\theta\qbin{1-h}{1-g}(v).
\end{equation}

\noindent This partial summation shows that the partition function
has a second order zero for $v\rightarrow 0$ from the left-moving
sector. This can be traced to the presence of two left-moving
massless gravitini in the untwisted sector and indicates \emph{at
least} $N=2$ space-time supersymmetry from this sector. However to
see that the \emph{full} $N=4+4$ supersymmetry is present we need to
show that an extra double zero is present in the partition function.
Then we are left to compute:
\begin{equation}
\mathcal{S}=\frac{1}{2}\sum_{h,g} (-1)^{\gamma g+\delta
h}\theta\qbin{1+h}{1+g}~\theta\qbin{1-h}
{1-g}~\theta\qbin{\gamma+h}{\delta+g}~\theta\qbin{\gamma-h}{\delta-g}.
\end{equation}
Defining
\begin{equation}
(A,B)=(1-h,1-g);~~ ~~(\gamma,\delta) =(1+H,1+G)~,
\end{equation}
and using the Jacobi identity associated to $(A,B)$, one can show
that:
$$
\mathcal{S}=(-1)^{GH+G}~\frac{1}{2}\sum_{A,B}(-1)^{A+B}~
\theta\qbin{A}{B}^2(v)~
\theta\qbin{A+H}{B+G}(v)~\theta\qbin{A-H}{B-G}(v)
$$
$$
=(-1)^{G(H+1)}~\theta\qbin{1}{1}^2(v)~\theta\qbin{1+H}{1+G}(v)~
\theta\qbin{1-H}{1-G}(v)
$$
\begin{equation}
=\theta\qbin{1}{1}^2(v)~\theta\qbin{\gamma}{\delta}^2(v),
\end{equation}
so that overall
\begin{equation}
Z_b=\frac{1}{\left|\eta\right|^{8}}~\theta\qbin{1}{1}^4~
\ov{\theta}\qbin{1}{1}^4 ~{1\over
2}\sum_{\gamma,\delta}~\theta\qbin{\gamma}{\delta}^6~
\ov{\theta}\qbin{\gamma}{\delta}^6~.
\end{equation}

\noindent Thus, we have  shown  explicitly that the $b-$twisted
partition function exhibits a zero of order four on the holomorphic
and anti-holomorphic sectors. The two extra zero's correspond to the
presence of two left-moving massless gravitini in the $b-$twisted
super-sector as we have mentioned above. Actually the computation of
the string helicity super-traces \cite{bk},\cite{Kiritsis:2007zz}
would let us conclude that the $\langle F,S,\ov{S}, b\rangle $ model
has a maximal $N=8$ supersymmetry.

\noindent
 Other choices of the GGSO projection
coefficients define non trivial lattice shifts. Choosing for instance
the ``factorized point'' of the $Z_{6,6}$  such that
\begin{equation}
Z_{6,6}\qbin{h}{g}_{\rm twisted}=Z_{4,4}\qbin{h}{g}_{\rm
twisted}~{\Gamma_{2,2}(T,U)\over |\eta|^4}
\end{equation}
where the twisted lattice $Z_{4,4}\qbin{h}{g}_{\rm twisted}$
correspond to the contribution of the
$\partial\phi^{I}=y^I\omega^I,~I=3,4,5,6 $ directions which are
twisted by $Z_2$. The $\Gamma_{2,2}(T,U)$ lattice is the
contribution of the untwisted directions $\partial\phi^{1,2}=(y
\omega)^{1,2}$. The latter depends on the $T,U$ moduli which are
associated to the 2-torus. In the fermionic construction $T$ and $U$
are fixed to the self-dual point $T=U=i$.

\noindent A way to break the right-moving space-time supersymmetry
``spontaneously" is to correlate the right-moving helicity with a
lattice shift. However, in order to preserve the bosonic content of
the massless spectrum it is necessary to keep some of the twisted
Ramond-Ramond states. This leads us to correlate the helicity
characters $(\ov{a},\ov{b})$, the twisted  $(h,g)$ characters with a
lattice shift of the $\Gamma_{2,2}$ lattice \cite{Rohm:1983aq}.

\noindent To do that one defines the shifted lattice sum as
$$
\Gamma_{2,2}\qbin{\ov{a}+h}{\ov{b}+g}=\sum_{n_i,m_j}
(-1)^{(\ov{a}+h)m_1+(\ov{b}+g)n_1+m_1n_1}
$$
\begin{equation}
\times\frac{T_2}{\tau_2}\exp \left[ -2i\pi B_{ij}m^in^j- \pi
G_{ij}\frac{(m^i+n^i\tau)(m^j+n^j\ov{\tau})}{\mathrm{Im}\,\tau}\right]
\end{equation}
with
\begin{equation}
G_{ij}={ {\rm Im} \,T  \over {\rm Im}\, U} ~\left[^{~1~~~~~~~{\rm
Re}\,U}_{{\rm Re}\, U~~~~~|U|^2} \right],~~~~
B_{ij}=\epsilon_{ij}~{\rm Re}\, T
\end{equation}
written in the Poisson dual form. This lattice sum differs from
$\Gamma_{2,2}(T,U)$ by the introduction of the modular invariant
phase $(-1)^{(\ov{a}+h)m_1+(\ov{b}+g)n_1+m_1n_1}$. Note that the
right helicity shift has the usual form of a right-moving
``temperature'' coupling \cite{Rohm:1983aq} while the
$(-1)^{hm_1+gn_1}$ shift acts on the twisted sectors
$(h,g)$ of the $Z_2$ orbifold. This phase modification does not change the modular
covariance properties of the lattice sum. Its effect is to make
massive the sectors corresponding to $\ov{a}+h=1~ mod~ 2$.

\noindent Some comments are in order :
\begin{itemize}
\item In the $\hat{\emptyset}$ super-sector we loose the R-R and NS-R sectors
which contain 8 vectors, 16 scalars, 12 spin $1/2$ fermions and 4
gravitini coming from the right moving side.
\item The $\hat{b}$ super-sector contains the NS-R and R-R sectors which provide
us with 8 vectors, 16 scalars and 16 spin $1/2$ fermions.
\end{itemize}

\noindent We see that overall, all these operations have had no
effect on the bosonic content of the theory with respect to the
$N=2+4$ model corresponding to the $b'$-orbifold. However, from the
fermionic fields point of view we have lost 4 gravitini but gained 4
additional spin $1/2$ fermions. This is what we call a $(3/2)\leftrightarrow(1/2)$
twist.

\noindent Given the number of vector fields it is clear that no
hypermultiplet is present in the spectrum. However since 16 among
the 30 scalar fields now come from the $b$-twisted sector it is not
immediate to conclude what is the special K\"ahler manifold that is
associated to the scalars. Several possibilities exist with this
dimension namely
\begin{equation}
{SU(1,15)\over U(1)\times SU(15)};~~~~{SO(2,14)\over{ SO(2)\times
SO(14)}} \times {SU(1,1)\over U(1)};~~~~{SO^*(12)~\over U(6)}
\end{equation}
The first two need either a rank 15 or rank 9 symmetry group to be
realized in a linear way  which is a too large  symmetry to be
realized in the model under consideration. On the other hand, one
can realize explicitly a rank 6 symmetry group through
\begin{equation}
SO(2)_{\chi^{1,2}}\times U(1)_{\phi,a}\times SU(2)^+_{y^{3,4,5,6}}
\times SU(4)_{\ov{\chi}^{3,4,5,6}}\subset U(6)
\end{equation}
where, in the above equation, the fields associated to each group
factor are indicated as lower indices.

\noindent Furthermore, from the coset decomposition of
\begin{equation}
{SO^*(12)~\over U(6)}~\rightarrow ~\left({SU(1,1)\over U(1)}\times
{SO(2,6)\over SO(2)\times SO(6)}\right)_{h=0} \times
\left({SU(2,4)\over U(1)\times SU(2)\times  SU(4)}\right)_{h=1},
\end{equation}
we recognize the coset structure of the 14 untwisted moduli from the
$\hat{\emptyset}$ super-sector ($h=0$) and the 16 RR-moduli coming
from the twisted $\hat b$-super-sector ($h=1$). This decomposition is
identical to the one of $N=2+6$ model described in ref.\cite{fk}.

\noindent We shall momentarily proceed to the construction of the other four
dimensional magic models, but we would like to make some comments on the
reduction and oxidation of ${\rm \cal M}^{N=2}_6$ to three and five
space-time dimensions:

\noindent i) The three dimensional case is obtained via $S^1$
compactification. The sets $S,{\ov S}$ and $b$ are taken to be the
same as in the four dimensional construction. In three dimensions
however the dimension of the scalar manifold is extended via 3d
duality transformation of the vector gauge bosons to scalars. In the
untwisted $h=0$ sector the dimension of the 3d scalar manifold
becomes: \\
{\bf 14} (4d-scalars)+ {\bf 14} (4d-vectors)+ {\bf 2}
(4d-graviphoton)+ {\bf 2} (3d-graviphoton, $g_{3,\mu}$)= {\bf 32}.\\
Altogether parameterize (via the $c-$map) the quaternionic manifold,
\begin{equation}
{\rm \cal M}^{D=3}_{h=0}={SO(4,8)\over SU(2)\times SU(2)\times
SO(8)}
\end{equation}
\noindent In the $h=1$ sector the dimension of the 3d scalar
manifold becomes:\\
{\bf 16} (4d-RR scalars) + {\bf 16} (4d-RR vectors) = {\bf 32}.\\
The 32 scalars from the $h=0$ sector together with the 32 scalars
from $h=1$ sector parameterize in 3d the quaternionic manifold of
the magic square \cite{frt,gst}:
\begin{equation}
{\rm \cal M}^{D=3}_6={E_{7(-5)}\over SU(2) \times SO(12)}~,
\end{equation}
as expected by a $c-$map operation on the four dimensional ${\rm
\cal M}^{N=2}_6$ magic model.

\noindent ii) The five dimensional case is obtained from ${\cal
M}^{N=2}_6$ magic model via one dimensional oxidation. Here also the
sets $S, {\ov S}$ and $b$ are the same as in the four dimensional
case. The only difference is the replacement of the two dimensional
lattice $\Gamma_{2,2}$ by the one dimensional lattice
$\Gamma_{1,1}$. Also, we identify $\chi^1\equiv \psi^5$ and
$y^1w^1\equiv \partial X^5,~{\ov y}^1{\ov w}^1\equiv {\ov \partial}
X^5$; $X^5$ is taken non-compact. In five dimensions the number of
scalars is reduced since they are becoming the 5th components of
higher spin fields. In the $h=0$ sector the 6 scalars parameterize
the manifold
\begin{equation}
{\rm \cal M}^{D=5}_{h=0}= SO(1,1)\times {SO(1,5)\over SO(5)} \equiv
SO(1,1)\times {SO(1,5)\over SP(4)}
\end{equation}
In the $h=1$ sector the RR-states are decomposed in 5d vectors and
5d scalars. The $1\over 2$ of the 4d-scalar degrees of freedom in
this sector are eaten by the 5d-vectors. So we are left with 8
5d-scalars that parameterize the manifold
\begin{equation}
{\rm \cal M}^{D=5}_{h=1}= {SP(2,4)\over SP(2)\times SP(4)}
\end{equation}
The 6 scalars from $h=0$ sector together with the $8$ from the $h=1$
sector parameterize in five space-time dimensions the manifold of
the magic square \cite{frt,gst},
\begin{equation}
{\rm \cal M}^{D=5}_6={SU^*(6)\over USP(6)}~,
\end{equation}
as expected by the supersymmetry conserving operations via {\it
Oxidation $\leftrightarrow$ Reduction}.

\section{Superstring construction  of the magic ${\rm \cal M}^{N=2}_3$}
\label{s:magicsquareN=3}

\noindent The other $N=2$ ``magic" theory we would like to construct
in four space-time dimensions is the hyper-free theory which has the
bosonic spectrum of the $N=3$ supergravity coupled to three extra
vector multiplets. The scalar manifold is K\"ahler and contains 18
scalars.
\begin{equation}
{\rm {\cal M}}_3={ SU(3,3)\over S( U(3) \times U(3))}
\end{equation}
Here also the string construction has to be asymmetric involving
asymmetric twists and lattice shifts.

\noindent Our starting point is a twisted realization of the $N=8$
based on the holomorphic and  anti-holomorphic basis sets
$$
b'_1 = \{\psi^\mu,\chi^{1,2},y^{3,4},y^{5,6} \}
$$
$$
\ov{b}'_1 =\{\ov{\psi}^\mu,\ov{\chi}^{1,2},\ov{y}^{3,4},\ov{y}^{5,6}
\}
$$
\begin{equation}
\ov{b}'_2=\{\ov{\psi}^\mu,\ov{\chi}^{3,4},\ov{y}^{1,2},\ov{y}^{5,6}\}.
\end{equation}
The holomorphic set  $A$
\begin{equation}
A=\{ y^{3,4}y^{5,6}w^{3,4}w^{5,6}~\}
\end{equation}
 will be  used  in our construction as well.
$b'_1$ induces a $Z^2_{b'_1}$ projection which seems to break the
left-moving supersymmetry from 4 to 2. Also  $\ov{b}'_1,\ov{b}'_2$
induce $Z^2_{\ov{b}'_1}\times Z^2_{\ov{b}'_2}$ projections that seem
to break the right-moving supersymmetry from 4 to 1. However, these
supersymmetry breakings are not efficient in general due to the
(anti-) holomorphic structure of the basis sets that imply the
presence of extra gravitini in the twisted sectors of the theory.
Thus, there is a choice of the GGSO coefficients where the
supersymmetry is maximal. Explicitly, this choice defines the
following partition function for the $N=(2+2)+(1+1+1+1)$ model
$\{S,b'_1;\ov{S},\ov{b}'_1,\ov{b}'_2; A\}$

$$
Z_{N=8}~=~\frac{1}{|\eta|^{24}}~~\frac{1}{2}\sum_{h_1,g_1,}
~\frac{1}{2}\sum_{\ov{h}_1,\ov{g}_1,}
~\frac{1}{2}\sum_{\ov{h}_2,\ov{g}_2}
~\frac{1}{2}\sum_{\gamma,\delta}~\frac{1}{2}\sum_{A,B}
$$

$$
\times ~~\frac{1}{2}\sum_{a,b}(-1)^{a+b+ab}~~
\theta\qbin{a}{b}~~\theta\qbin{a}{b}~~\theta\qbin{a+{h_1}}{b+{g_1}}~~
\theta\qbin{a-{h_1}}{b-{g_1}}~~(-)^{h_1g_1+AB}
$$

$$
\times ~~~~\theta\qbin{\gamma}{\delta}\theta\qbin{\gamma}{\delta}
~~\theta\qbin{\gamma-A}{\delta-B}\theta\qbin{\gamma+A+{h_1}}{\delta+B+{g_1}}~~
\theta\qbin{\gamma-A}{\delta-B}\theta\qbin{\gamma+A-{h_1}}{\delta+B-{g_1}}~~~~
$$

$$
\times~~\ov{\theta}\qbin{{\gamma}}{{\delta}}
\ov{\theta}\qbin{{\gamma}+{\ov{h}_1}}{{\delta}+{\ov{g}_1}}~~
\ov{\theta}\qbin{{\gamma}}{{\delta}}\ov{\theta}\qbin{{\gamma}-{\ov{h}_2}}{{\delta}-
{\ov{g}_2}}
~~\ov{\theta}\qbin{{\gamma}}{{\delta}}
\ov{\theta}\qbin{{\gamma}-{\ov{h}_1+\ov{h}_2}}{{\delta}-{\ov{g}_1+\ov{g}_2}}
~(-)^{\ov{h}_1\ov{g}_1+\ov{h}_2\ov{g}_2}
$$

\begin{equation}
\times~~
\frac{1}{2}\sum_{\ov{a},\ov{b}}(-1)^{\ov{a}+\ov{b}+\ov{a}\ov{b}}~~
\ov{\theta}\qbin{\ov{a}}{\ov{b}}~~\ov{\theta}\qbin{\ov{a}+{\ov{h}_1}}
{\ov{b}+{\ov{g}_1}}~~
\ov{\theta}\qbin{\ov{a}+{\ov{h}_2}}{\ov{b}+{\ov{g}_2}}~~
\ov{\theta}\qbin{\ov{a}-{\ov{h}_1-\ov{h}_2}}{\ov{b}-{\ov{g}_1-\ov{g}_2}}
\end{equation}

\noindent The choice of the phases and the arguments of the
$\theta$-functions is dictated by modular invariance and the
existence of maximal supersymmetry. Indeed, the existence of 4
left-moving supersymmetries can be shown explicitly by using the
Jacobi identity associated to the the arguments $(a,b)$ and then to
$(h_1, g_1)$ as previously. The 4 right-moving supersymmetries can be
visualized by using first the Jacobi identity associated to
$(\ov{a},\ov{b})$ and then the one associated to
$(\ov{h}_1,\ov{g}_1)$ and then to $(\ov{h}_2,\ov{g}_2)$.

 \noindent
The massless spectrum of the above  $N=8$-twisted construction
is the following:\\

 \noindent
\underline{$\hat{\emptyset}$ super-sector} ~
$$
\left[ (\psi^\mu,\chi^{1,2}) \, \oplus\, sp(\psi^\mu,\chi^{1,2})_-
sp(\chi^{3,4,5,6})_+ \right]
$$
\begin{equation}
\otimes \left[ \ov{\psi}^\mu \, \oplus\,
sp(\ov{\psi}^\mu)_-sp(\ov{\chi}^{1,2})_-sp(\ov{\chi}^{3,4})_+
sp(\ov{\chi}^{5,6})_+ \right],
\end{equation}

\noindent \underline{$\hat{b}'_1$ super-sector} ~
$$
\left[ (sp(\psi^\mu,\chi^{1,2})_-  sp(y^{3,4,5,6})_+ \,
\oplus\,sp(\chi^{3,4,5,6})_+ sp(y^{3,4,5,6})_+ \right]
$$
\begin{equation}
\otimes \left[ (\ov{\psi}^\mu,\ov{\chi}^{1,2}) \, \oplus\,
sp(\ov{\psi}^\mu)_-sp(\ov{\chi}^{1,2})_-sp(\ov{\chi}^{3,4})_+
sp(\ov{\chi}^{5,6})_+ \right],
\end{equation}

\noindent \underline{$\hat{\ov{b}'_1}$ super-sector} ~
$$
\left[ (\psi^\mu,\chi^{1,2}) \, \oplus\, sp(\psi^\mu,\chi^{1,2})_-
sp(\chi^{3,4,5,6})_+ \right]
$$
\begin{equation}
 \otimes \left[ \left( sp(\ov{\psi}^\mu)_-sp(\ov{\chi}^{1,2})_+
 \, \oplus\,
 sp(\ov{\chi}^{3,4})_+
sp(\ov{\chi}^{5,6})_+\right)\oplus sp(\ov{y}^{3,4})_+
 sp(\ov{y}^{5,6})_+ \right],
\end{equation}

\noindent \underline{$\hat{\ov{b}'_2}$ super-sector} ~
$$
\left[ (\psi^\mu,\chi^{1,2}) \, \oplus\, sp(\psi^\mu,\chi^{1,2})_-
sp(\chi^{3,4,5,6})_+ \right]
$$
\begin{equation}
 \otimes \left[ \left( sp(\ov{\psi}^\mu)_-sp(\ov{\chi}^{3,4})_+
 \, \oplus\,
 sp(\ov{\chi}^{1,2})_+
sp(\ov{\chi}^{5,6})_+\right)\oplus sp(\ov{y}^{1,2})_+
 sp(\ov{y}^{5,6})_+ \right],
\end{equation}

\noindent \underline{$\hat{\ov{b}'_1}\hat{\ov{b}'_2}$ super-sector} ~
$$
\left[ (\psi^\mu,\chi^{1,2}) \, \oplus\, sp(\psi^\mu,\chi^{1,2})_-
sp(\chi^{3,4,5,6})_+ \right]
$$
\begin{equation}
 \otimes \left[ \left( sp(\ov{\psi}^\mu)_-sp(\ov{\chi}^{5,6})_+
 \, \oplus\,
 sp(\ov{\chi}^{1,2})_+
sp(\ov{\chi}^{3,4})_+\right)\oplus sp(\ov{y}^{1,2})_+
 sp(\ov{y}^{3,4})_+ \right],
\end{equation}

\noindent \underline{$\hat{b}'_1\hat{\ov{b}'_1}$ super-sector} ~
$$
\left[ (sp(\psi^\mu,\chi^{1,2})_-  sp(y^{3,4,5,6})_+ \,
\oplus\,sp(\chi^{3,4,5,6})_+ sp(y^{3,4,5,6})_+ \right]
$$
\begin{equation}
 \otimes \left[ \left( sp(\ov{\psi}^\mu)_-sp(\ov{\chi}^{1,2})_+
 \, \oplus\,
 sp(\ov{\chi}^{3,4})_+
sp(\ov{\chi}^{5,6})_+\right)\oplus sp(\ov{y}^{3,4})_+
 sp(\ov{y}^{5,6})_+ \right],
\end{equation}

\noindent \underline{$\hat{b}'_1\hat{\ov{b}'_2}$ super-sector} ~
$$
\left[ (sp(\psi^\mu,\chi^{1,2})_-  sp(y^{3,4,5,6})_+ \,
\oplus\,sp(\chi^{3,4,5,6})_+ sp(y^{3,4,5,6})_+ \right]
$$
\begin{equation}
 \otimes \left[ \left( sp(\ov{\psi}^\mu)_-sp(\ov{\chi}^{3,4})_+
 \, \oplus\,
 sp(\ov{\chi}^{1,2})_+
sp(\ov{\chi}^{5,6})_+\right)\oplus sp(\ov{y}^{1,2})_+
 sp(\ov{y}^{5,6})_+ \right],
\end{equation}

\noindent \underline{$\hat{b}'_1\hat{\ov{b}'_1}\hat{\ov{b}'_2}$
super-sector} ~
$$
\left[ (sp(\psi^\mu,\chi^{1,2})_-  sp(y^{3,4,5,6})_+ \,
\oplus\,sp(\chi^{3,4,5,6})_+ sp(y^{3,4,5,6})_+ \right]
$$
\begin{equation}
 \otimes \left[ \left( sp(\ov{\psi}^\mu)_-sp(\ov{\chi}^{5,6})_+
 \, \oplus\,
 sp(\ov{\chi}^{1,2})_+
sp(\ov{\chi}^{3,4})_+\right)\oplus sp(\ov{y}^{1,2})_+
 sp(\ov{y}^{3,4})_+ \right],
\end{equation}

\noindent Even though the untwisted {$\hat{\emptyset}$ super-sector}
corresponds to a theory with $N=2+1$ supersymmetry, the
supersymmetry is extended to $N=(2+2)+(1+3)$ because of the extra
left- and right-moving  gravitini arising from  the $\hat{b}'_1$,
$\hat{\ov b}'_1$,$\hat{\ov b}'_2$ and $\hat{\ov b}'_1 \hat{\ov b}'_2$
twisted super-sectors. There are two extra left-moving gravitini from
$\hat{\ov b}'_1$ and three right moving ones from $\hat{\ov
b}'_1$,$\hat{\ov b}'_2$ and $\hat{\ov b}'_1 \hat{\ov b}'_2$. This
supersymmetric extensions happens due to the left-and right- moving
``holomorphic'' structure of $b'_1,{\ov b}'_2,{\ov b}'_3$. Thus, the
twisted construction still has maximal $N=8$ supersymmetry
constructed in a twisted manner.

\noindent A way to reduce the left- and right- supersymmetry is to
couple the lattice characters $(\gamma,\delta)$ and $(A,B)$ to the
left- and right- helicities and (twisted) $R$-symmetry charges. We
will first construct two versions of the $N=2+(1+1)$ supergravity
model containing six and eight extra vector multiplets. Then we will
reduce further the supersymmetry to obtain the magic ${\rm \cal
M}^{N=2}_3$ .

\subsection{$N=2+(1+1)$ supergravity, with
${\rm \cal M}^{N=4}_{n_A}={SU(1,1)\over U(1)}\times{SO(6,~n_A)\over
S(6)\times SO(n_A)}$ }

\noindent In order to reduce the supersymmetry to $N=2+(1+1)$ one
has to eliminate from the massless sectors the gravitini coming from
the $\hat{b}'_1$, $\hat{\ov b}'_1$ and $\hat{\ov b}'_2$ twisted
super-sectors. One way to do that is to use the lattice characters
$(\gamma,\delta)$ and $(A,B)$ to impose $Z_2$ projections. Inserting
in the $N=8$ partition function $Z_{N=8}$ the phase
$$
Z_{N=8}~~\longrightarrow~~ Z_{N=4}^A
$$
\begin{equation}
=Z_{N=8}~(-)^{\delta(A+h_1+\ov{h}_1+\ov{h}_2)+
\gamma(B+g_1+\ov{g}_1+\ov{g}_2)+
(A+h_1+\ov{h}_1+\ov{h}_2)(B+g_1+\ov{g}_1+\ov{g}_2)}
\end{equation}
imposes in the massless states the constraint
\begin{equation}
(-)^{A+h_1+\ov{h}_1+\ov{h}_2}=+1
\end{equation}
which eliminates the $\hat{b}'_1$, $\hat{\ov b}'_1$, $\hat{\ov b}'_2$
as well as the $\hat{b}'_1\hat{\ov b}'_1\hat{\ov b}'_2$ super-sectors.
Naively one obtains a $N=2+2$ supergravity  model with two
left- one right-moving supersymmetries from the $\hat{\emptyset}$
super-sector  and one right-moving supersymmetry from the $\hat{\ov
b}'_1\hat{\ov b}'_2$ super-sector.

\noindent The same sectors can be eliminated with a different choice
of the phase,
$$
Z_{N=8}~~\longrightarrow~~ Z_{N=4}
$$
\begin{equation}
=Z_{N=8}~(-)^{\delta(h_1+\ov{h}_1+\ov{h}_2)+
\gamma(g_1+\ov{g}_1+\ov{g}_2)+
(h_1+\ov{h}_1+\ov{h}_2)(g_1+\ov{g}_1+\ov{g}_2)}
\end{equation}
imposing the constraint
\begin{equation}
(-)^{h_1+\ov{h}_1+\ov{h}_2}=+1
\end{equation}
Both $Z_{N=4}^A$ and $Z_{N=4}$ have  $N=4$ supersymmetry.
 However, due to the holomorphic
structure of the $A$-set in the $Z_{N=4}^A$ model, some extra
massless states arise from the $\hat{A}\hat{b}'_1\hat{\ov
b}'_1\hat{\ov b}'_2$ super-sector. In the $Z_{N=4}$ as well as in the
initial $Z_{N=8}$ model these extra massless states are projected
out due to the $B$-projection.

\noindent The massless states of the $Z_{N=4}$ are those of the $N=4$
supergravity coupled to six extra vector multiplets. There are
38 scalars parameterizing the manifold:
\begin{equation}
{\rm \cal M}^{N=4}_6={SU(1,1)\over U(1)}\times{SO(6,6)\over
SO(6)\times SO(6)}~.
\end{equation}

\noindent In the $Z_{N=4}^A$ construction there are extra massless states from
the $\hat{A}\hat{b}'_1\hat{\ov{b}'_1}\hat{\ov{b}'_2}$ super-sector

 \noindent \underline{$\hat{A} \hat{b}'_1 \hat{\ov{b}'_1}\hat{\ov{b}'_2}$
super-sector} ~
$$
\left[ (sp(\psi^\mu,\chi^{1,2})_-  sp(w^{3,4,5,6})_+ \,
\oplus\,sp(\chi^{3,4,5,6})_+ sp(w^{3,4,5,6})_+ \right]
$$
\begin{equation}
 \otimes \left[ \left( sp(\ov{\psi}^\mu)_-sp(\ov{\chi}^{5,6})_+
 \, \oplus\,
 sp(\ov{\chi}^{1,2})_+
sp(\ov{\chi}^{3,4})_+\right)\oplus sp(\ov{y}^{1,2})_+
 sp(\ov{y}^{3,4})_+ \right].
\end{equation}

\noindent In the $Z_{N=4}^{A}$ the extra vector multiplets are eight
and the total number of scalars is 50 that parameterize  the
manifold:
\begin{equation}
{\rm \cal M}^{N=4}_8={SU(1,1)\over U(1)}\times{SO(6,8)\over
SO(6)\times SO(8)}~.
\end{equation}
\noindent In the next subsection starting from the ${\rm \cal
M}^{N=4}_8$ model, we will construct the magic ${\rm \cal
M}^{N=2}_3$  by an ``helicity twisting mechanism''
similar to the one we had introduced in the construction of the
${\rm \cal M}^{N=2}_6$.

\subsection{${\rm \cal M}^{N=4}_8 \longrightarrow $ magic ${\rm \cal
M}^{N=2}_3$}

\noindent The magic ${\rm \cal M}^{N=2}_3$ can be obtained  from the
asymmetric $Z_{N=4}^{A}$  construction breaking the 2 right-moving
supersymmetry via the insertion of a phase
$$
Z_{N=4}^{A}~\longrightarrow~Z_{N=2}^{{\rm Magic}_3}~
$$
\begin{equation}=~Z_{N=4}^{A}~(-)^{B(h_1+\ov{a})+A(g_1+\ov{b})+AB}
\end{equation}
which couples the left-twisted arguments $(h_1,g_1)$ and  the
right-helicity charges $(\ov{a},\ov{b})$ with the lattice arguments
$(A,B)$. The phase factor $(-)^{AB}$ cancels the one appearing in
the $Z_{N=8}$ and $Z_{N=4}^{A}$ partition function. The induced
$B$-constraint for the massless states is
 $$(-)^{h_1+\ov{a}}=+1$$ eliminating  all massless states
of ${\rm \cal M}^{N=4}_8$ with $(-)^{h_1+\ov{a}}=-1$. The remaining
states are then:\\

\noindent\underline{\bf Massless spectrum of the magic ${\rm \cal
M}^{N=2}_3$}\\

 \noindent \underline{$\hat{\emptyset}$
super-sector, $(h_1=0,~\ov{a}=0)$} ~
\begin{equation}
\left[ (\psi^\mu,\chi^{1,2}) \, \oplus\, sp(\psi^\mu,\chi^{1,2})_-
sp(\chi^{3,4,5,6})_+ \right] \otimes \left[ \ov{\psi}^\mu \,
\right],
\end{equation}
1 graviton, 2 gravitini, 2 spin-1/2 fermions,
 2 gauge bosons and 2 scalars.\\

\noindent \underline{$\hat{\ov{b}'_1}\hat{\ov{b}'_2}$ super-sector,
$(h_1=0,~ \ov{a}=0)$} ~
$$
\left[ (\psi^\mu,\chi^{1,2}) \, \oplus\, sp(\psi^\mu,\chi^{1,2})_-
sp(\chi^{3,4,5,6})_+ \right]
$$
\begin{equation}
 \otimes \left[
 sp(\ov{\chi}^{1,2})_+
sp(\ov{\chi}^{3,4})_+sp(\ov{y}^{1,2})_+
 sp(\ov{y}^{3,4})_+ \right],
\end{equation}
4  spin-1/2 fermions, 2 gauge bosons and 4 scalars.\\

\noindent \underline{$\hat{b}'_1\hat{\ov{b}'_1}$ super-sector,
$(h_1=1,~\ov{a}=1)$} ~
$$
\left[ (sp(\psi^\mu,\chi^{1,2})_-  sp(y^{3,4,5,6})_+ \,
\oplus\,sp(\chi^{3,4,5,6})_+ sp(y^{3,4,5,6})_+ \right]
$$
\begin{equation}
 \otimes \left[ sp(\ov{\psi}^\mu)_-sp(\ov{\chi}^{1,2})_+
 sp(\ov{y}^{3,4})_+
 sp(\ov{y}^{5,6})_+ \right],
\end{equation}
4  spin-1/2 fermions, 2 gauge bosons and 4 scalars.\\

\noindent \underline{$\hat{b}'_1\hat{\ov{b}'_2}$ super-sector,
$(h_1=1,~\ov{a}=1)$}

$$
\left[ (sp(\psi^\mu,\chi^{1,2})_-  sp(y^{3,4,5,6})_+ \,
\oplus\,sp(\chi^{3,4,5,6})_+ sp(y^{3,4,5,6})_+ \right]
$$
\begin{equation}
 \otimes \left[sp(\ov{\psi}^\mu)_-sp(\ov{\chi}^{3,4})_+
sp(\ov{y}^{1,2})_+
 sp(\ov{y}^{5,6})_+ \right],
\end{equation}
4  spin-1/2 fermions, 2 gauge bosons and 4 scalars.\\

 \noindent
\underline{$\hat{A}\hat{b}'_1\hat{\ov{b}'_1}\hat{\ov{b}'_2}$ super-sector
$(h_1=1,~ \ov{a}=1)$}
$$
\left[ (sp(\psi^\mu,\chi^{1,2})_-  sp(w^{3,4,5,6})_+ \,
\oplus\,sp(\chi^{3,4,5,6})_+ sp(w^{3,4,5,6})_+ \right]
$$
\begin{equation}
 \otimes \left[sp(\ov{\psi}^\mu)_
 sp(\ov{\chi}^{5,6})_+sp(\ov{y}^{1,2})_+
 sp(\ov{y}^{3,4})_+ \right].
\end{equation}
4 spin-1/2 fermions, 2 gauge bosons and 4 scalars.\\

\noindent The $N=2$ graviton multiplet comes from the
$\hat{\emptyset}$ super-sector. The same super-sector contains one
vector multiplet as well. There are eight additional vector
multiplets from the other super-sectors. In total the number of
scalars is 18 and they parameterize the Magic $N=2$ manifold
\begin{equation}
{\rm {\cal M}}_3={ SU(3,3)\over S( U(3) \times U(3))},
\end{equation}
This manifold is based to the $N=2$ holomorphic prepotential
\begin{equation}
 F(Z_0,Z^{ij})=-i~{{\rm Det}( Z^{ij})\over Z_0}
 =-iZ_0^2~{{\rm Det} (t^{ij})}~,
 \end{equation}
\noindent where the $3\times 3$ matrix $Z^{ij}$ parameterizes the
nine complex scalars ($t^{ij}=Z^{ij}/Z_0$). The K\"ahler potential
associated to the magic ${\rm \cal M}^{N=2}_3$ is:
\begin{equation}
K=-\log~i{\rm Det}(t^{ij}-\ov{t}^{ij})
 \end{equation}
\noindent and has the property to be identical to the $N=3$
supergravity coupled to three extra vector multiplets.

\noindent Utilizing  the same basis sets as in the 4d construction
of ${\rm \cal M}^{N=2}_3$ and performing similar operations it is
straightforward to define the reduced theory in three space time
dimensions. Via 3d duality transformation acting on 3d vectors the
3d manifold is extended to:\\
{\bf 18} (4d-scalars) + {\bf 18}  (4d-vectors) + {\bf 2}
(4d-graviphoton) + {\bf 2}  (3d-graviphoton, $g_{3,\mu}$) = {\bf
40}.\\
The obtained 3d theory is that of the magic square with scalar
manifold
\begin{equation}
{\rm \cal M}^{D=3}_3={E_{6(2)}\over SU(2)\times SU(6)}~,
\end{equation}
as expected by the $c-$map operation.

\noindent One expects via {\it oxidation} to five space-time
dimensions to construct the  scalar manifold of the magic square
with 8 5d-scalars:
\begin{equation}
{\rm \cal M}^{D=5}_3 = {SL(3,C)\over SU(3)}~.
\end{equation}
Although this operation looks straightforward in field theory set-up
there is an obstruction in the above stringy construction where all
internal left-moving coordinates are twisted. It is therefore
impossible to construct this model in our set-up. This obstruction
however does not prevent the stringy existence of ${\rm \cal
M}^{D=5}_3$ via asymmetric orientifold construction or else which
appear non-perturbative from the type II ``close strings'' framework.
Hopefully, we will return and try to clarify this obstruction in
near future.

\section{Discussion}\label{s:conclusions}

\noindent Several type II superstring vacua  with $N=2$
supersymmetry can be constructed. In this work we focussed our
attention on those which do not contain in their massless
spectrum any hypermultiplet. In this class of vacua the scalar
manifold of the vector supermultiplets is always K\"ahler. It is
interesting that for all four dimensional hyper-free constructions, the
internal compactification is necessarily {\it not a Calabi-Yau
manifold} or more generally the world-sheet superconformal symmetry
is not based on $ {\rm \cal N}=(2,2)$
 but rather on ${\rm\cal N}=(4,1)$. Indeed, their constructions
is left-right asymmetric and can be realized  by asymmetric
orbifolds via 2d-fermionic construction.

\noindent The minimal hyper-free theory with only one massless
vector multiplet has been constructed. 
This theory contains a single {\it
minimally coupled vector multiplet $S$} associated to the
axion-dilaton pair. This theory is exotic from the viewpoint of
Calabi-Yau compactification where the vector multiplets are
conformaly (non-minimally) coupled. Furthermore, this theory is
universal in the sense that is a part of the spectrum of all the 
other more complex models.

\noindent Among the hyper-free $N=2$ theories two of them are
special. They  are known to be in correspondence with the symmetric
spaces of the ``magic square" and furthermore are associated to the
Jordan algebras $J_3^{\mathbb{C}}$ and $J_3^{\mathbb{H}}$. They have
the additional remarkable property to share the same scalar field
content as some supergravity models with more supersymmetry, $N=6$
and $N=3$. The one associated to the $N=6$ turns out to be very
special since not only the scalar degrees of freedom but all the
bosonic massless degrees of freedom are the same as the $N=6$
supergravity theory.

\noindent The superstring realization of the two $N=2$ of the magic
square turns out to be non-trivial. We were able to construct them
by introducing a ``twisting mechanism" that eliminates the extra
gravitini of the $N=3$ and $N=6$ supergravities and creates at the
same time the extra spin-${1\over 2}$ fermions and spin-1 gauge
bosons that are necessary to balance the $N=2$ boson-fermion
degeneracy.

\noindent The ``twisting mechanism" is interesting by itself. It is
a well defined operation in string theory and is based on
``holomorphic $Z_2$-orbifold stringy constructions":\\
i) The four left-moving gravitini are reduced to two as usually by
a $Z_2$-projection.\\
ii) Due to the {\it holomorphic structure} of the projection {\it
two extra gravitini} appears in the ``twisted" sector and thus
obtain a ``twisted $N=(2+2)+4$ realization" of the $N=8$ supergravity.\\
iii) The breaking of the four right-moving supersymmetry is realized
via a coupling of the lattice charges to the right-helicity charge
$\ov{a}$ and left-twisted charge $h$, imposing the constrain
$(h+\ov{a})=0~ mod~2$. This constrain breaks the two left- and the four right-moving
supersymmetry but keeps the gauge bosons and scalars coming from the
$N=2+4$ Ramond-Ramond states.

\noindent The above three steps define the ``twisting mechanism"
applied in the case of the magic $N=2$ associated to the $N=6$
supergravity. Although this mechanism is well defined  in string
theory it is not yet known how it could be realized in a $Z_2$-truncated
supergravity for two main reasons:\\
 i) From where could the twisted
gravitini appear?\\
 ii) How one can define the $R$-symmetry
charges associated to the ``twisted states"?

\noindent The ``twisting mechanism" is even more involved in the
case of the $N=2$ magic associated to the $N=3$ supergravity coupled
to three extra  vector multiplets. Here one starts from a ``twisted
$N=2+(1+1)$ realization" of $N=4$ supergravity coupled to eight extra
vector multiplets. Then, the constraint  $(h+\ov{a})=0~ mod~2$ is
introduced which reduces the supersymmetry to $N=2+0$. Here also the
construction is ``stringy". It is an open interesting problem if an
analogous construction can be realized in a $Z_2^n$-truncated
supergravity theories.

\noindent By reduction to three space-time dimensions, we are able
to construct at the string level two other theories of the magic
square namely: ${\rm \cal M}^{D=3}_6={E_{7(-5)}\over SU(2) \times
SO(12)}~,$ and $~{\rm \cal M}^{D=3}_3={E_{6(2)}\over SU(2)\times
SU(6)}$.

\noindent By oxidation in five space-time dimensions the
construction of ${\rm \cal M}^{D=5}_6={SU^*(6)\over USP(6)}~$ of the
magic square is also achieved at the string level. However, in our
stringy set-up, there is an obstruction to define ${\rm \cal
M}^{D=5}_3 = {SL(3,C)\over SU(3)}~$  since all six right-moving
coordinates are twisted  and this prevents the oxidation procedure in
five-dimensions. On the other hand this obstruction does not implies
in general the non-existence at the string level of ${\rm \cal
M}^{D=5}_3$. It is possible that this theory exists at the string
level via other constructions like ``asymmetric orientifolds" that
may appear as non-petrurbative constructions from ``closed string"
set-up we had explored here. It remains an open and very interesting
question the stringy existence of the other theories of the magic
square, in three, four and five space-time dimensions:
$$
D=5~~~~~~~{SL(3,R)\over SO(3)}~[5],~~~~~~~~~~~~~~{SL(3,C)\over
SU(3)}~[8],~~~~~~~~~~~~{SU^*(6)\over
USP(6)}~[14],~~~~~~~~~~~~~~{E_{6(-26)}\over F_4}~[26]
$$
$$
D=4~~~~~{SP(6,R)\over U(3)}~[12],~~~~~{SU(3,3)\over U(1)\times
SU(3)\times
  SU(3)}~[18],
~~~~{SO^*(12)\over U(6)}~[30], ~~~~~~~~{E_{7(-25)}\over U(1)\times
E_6}~[54]
$$
$$
D=3~~{F_{4,4}\over USP(6)\times SU(2)}~[28],~~~{E_{6(2)}\over
SU(2)\times SU(6)}~[40], ~~~{E_{7(-5)}\over SO(12)\times
SU(2)}~[64],~~~ {E_{8(-24)}\over SU(2)\times E_7}~[112]
$$

\noindent Indeed, the explicit constructions at the string level of
the magic $N=2$ theories extends the validity at the string level
of the entropy formulas obtained via the $BPS$ and non-$BPS$ attractor
mechanism introduced in refs\cite{Ferrara:1997tw}.

\noindent Finally, it will be very interesting to study
supersymmetric string vacua and their effective supergravity theories
that will be eventually obtained via a ``generalized twisting
mechanism";  not only in type II theories,  but also in heterotic as
well as in type II orientifolds with brane and fluxes.

\noindent After the submission of this work an interesting
paper appeared, by M. Bianchi and S. Ferrara \cite{BianchiFerrara}, where
two other magic square $N_4=2$ theories are obtained via an asymmetric
orientifold construction, namely $$\rm{ \cal M}_7={E_{7(-25)}\over U(1)\times
E_6}$$ in four space-time dimensions and
 $$\rm{ \cal M}_8={E_{8(-24)}\over SU(2)\times E_7}$$
in three dimensions. Even more interesting is the simultaneous appearance 
in their  construction of the two magic scalar manifolds in the same
four dimensional $N=2$ theory, with $\rm{ \cal M}_7$ as the scalar manifold
of the vector multiplets and  $\rm{ \cal M}_8$ the one of the
hypermultiplets. In three space time dimensions one obtains a {\it double 
magic theory} based on  $\rm{ \cal M}_8\times \rm{ \cal M}_8$, since  
$\rm{ \cal M}_8$ is derived by  $\rm{ \cal M}_7$ via a three dimensional
c-map. This doubling of the manifold is absent in our hyper-free
construction. In the same work M. Bianchi and S. Ferrara underlined   
the importance of string magic theories for the validity of the
entropy formulae obtained via the $BPS$ and non-$BPS$ attractor
mechanism, (for an updated view of the subject, see \cite{BellucciFerrara}).

\section*{Acknowledgments}

\noindent We are grateful to Sergio Ferrara and Massimo Bianchi for useful
discussions.\\
\noindent
This work is  partially supported by the EU
under the contracts MRTN-CT-2004-005104, MRTN-CT-2004-512194,
MRTN-CT-2004-503369, MEXT-CT-2003-509661, CNRS PICS 2530, 3059 and
3747, ANR (CNRS-USAR) contract  05-BLAN-0079-01.


\end{document}